\documentstyle[preprint,aps,epsf]{revtex}
 
\tightenlines

\begin{document}

\title{
Hybrid Exciton State in Quantum Dot- Dendrite System: The Green Functions}

\author{
{\bf
Nguyen Que Huong and Joseph L. Birman} \\
Physics Department, The City College, CUNY\\
Convent Ave. at 138 St, New York, N.Y. 10031, 
USA} 
                                             
\date{\today}
\maketitle

\begin{abstract}

A model is proposed to study the hybrid exciton in a quantum dot-dendrimer systems. The
semiconductor organic hybrid exciton
 is studied using a "real space" Green's function method and a diagrammatic
technique. The energy of the hybrid
 exciton as well as the Green function matrix elements have been
calculated for different quantum dot-dendrimer
 systems, and the method can be applied for systems with different
structures. Using the double-time Green's
 functions the optical processes can be calculated. The optical properties
of the systems are controllable by the size
 and structure of the QD-dendrimer systems. 

\end{abstract}
\newpage
{\bf I.Introduction.}
\vspace{0.5cm}

With the development of nano- and biotechnology, at present dendrimers
and dendrimeric systems are being studied intensively and many systems have become 
available[1-6].
Dendrimers are nanosize, highly symmetric and perfectly hyperbranched macromolecules with
controlled structure, which are good candidates to behave as optimal energy funnels. Dendrimers
with controlable sizes and structure can be used as light emmiters and can serve as 
building blocks in nano-devices [3].

Recent studies support the interest in preparing systems with the combination of properties of
semiconductor nanocrystal and organic dendrimer. Semiconductor nanocrystals, or quantum dots, 
which are semiconductor spheres with radius around $30\dot{A}-100\dot{A}$, have very strong 
quantum confinement effects. 
These quantum size effects of the motion and behavior of the carriers like 
electrons and holes lead to many interesting optical and structural phenomena[7-8].The optical 
properties and optical proceeses of a quantum dot strongly depend on the dot size and therefore 
can be controlled by the size of the nanocrystals.
The fact that  the electronic states in the quantum dot are widely separated due to quantum confinement, 
and the emission region is controllable by changing the dot size makes it interesting to use the 
dot as a photoactive element at the center of the dendrimer. The emision can be tuned by changing the 
dot size as well as the structure of the dendrimer[4].

In a solid, excitons play a fundamental role in optical properties, for example by determining 
the optical processes 
happening close to the band gap. There are two kinds of excitons in solids. The Wannier Mott- 
exciton in semiconductors is relatively weakly bound with the Bohr exciton radius $a_B \sim 30 
A-100A$ and therefore is called the large-radius 
exciton. 
The Frenkel exciton is the electronic state of a molecular or strongly ionic crystal, 
strongly localized with the exciton radius $a \sim 5A$.
But while the oscillator strength of the 
Wannier-Mott exciton is weak, the oscillator strength of the Frenkel exciton is very strong.
Because of the difference between these two kind of exciton it would be interesting to create some 
system with an hybrid excitation state which has the composite properties of both kinds of 
exciton.
Then one can expect to have an hybrid exciton with very strong oscillator strength as well 
as very large exciton radius.
Several specific systems of this semiconductor-organic combination  were 
proposed[9-12] such as a neighboring structure of organic and semiconductor quantum wells[9], neighboring 
organic and inorganic quantum wires[10], a single quantum dot in an organic shell[11] and the system 
of quantum dot array embedded in an organic host[12]. The obtained hybrid exciton also has been 
claimed to have very large non-linearities.

In this paper, we consider the hybrid exciton formed in the system with a quantum dot at the core 
of a dendrimer. In the quantum dot-dendrimer system, the Wannier exciton in the dot 
will interact with the Frenkel exciton in the dendrimer through dipole-dipole interaction to form 
some hybrid excitation. This hybrid exciton will play an important role in  optical properties 
of the system and by changing the size and structure of the nanocrystal-dendrimer, one can 
expect to change the hybrid exciton properties and therefore to control optical processes of the 
system such as by tuning emmision region or to change the nonlinear  susceptibilities of the 
system.    

We will use the Green's function approach to solve the problem. The double-time Green's 
functions in the real space with the diagram technique is very useful and important in 
solving the problems with complicated structures. The Green's functions are directly 
related to properties of the system, so once the Green's functions are  calculated, the 
optical properties such as non-linearities, fluorescence, scattering etc. can be evaluated. 

\vspace{0.5cm}

 {\bf II. The Model for the Quantum Dot-Dendrite system }

\vspace{0.5cm}

We consider a model of a quantum dot as a spherical core at the center, which is attached
to three dendrimeric branches.
Each dedrimeric branch consists of tertiary amine groups linked
by three-carbon chains.
Then each monomer unit (branching point) is attached to two protein branches,and so on (Fig.1).
There is coupling between the quantum dot and the attached protein branches,
between  molecules in the same branches, as well as between the molecule at
the branching point and the protein chain attached to it. But
no coupling between molecules of different branches is assumed.

The Wannier exciton in the quantum dot and the Frenkel exciton in each organic
molecule interact with each other through dipole-dipole interaction.
Here we assume only nearest neigbor interaction.

The tight-binding Hamiltonian of the system  can be written as the following:

\begin{equation}
H = \sum_{i=1}^{N}{\cal E}_{i}a^{+}_ia_i +\sum_{i,j}V_{ij}a^{+}_ia_j
\end{equation}
\noindent

Here $a^+_i$, $a_i$ are exciton creation and annihilation operators,
${\cal E}_{i}$ is the exciton energy at each site, with i labelling the sites
of quantum dot and molecules at the end of each generation of the dendrimeric
branch. For the quantum dot the energy will be the energy of the exciton confined in the dot[8]
and for the molecular sites it will be the Frenkel exciton energy.  
$V_{ij}$ is the interaction integral between excitons in 
different sites, or in other words, between exciton of different dendrimer generations.
Note  that by $i$ and $j$ here we  number sites in a "macroscopic" scale. 
Actually, at this scale $i, j$ are the sites of
the molecules at the branching point, i.e. the end of each generation.
For example, the quantum dot is located
at $i=0$, the molecules of the monomer unit at the end of the first generation located
at the site $i=1$, etc...
The molecules within each chain linking different
generations are counted in the "microscopic scale".
The interaction
between molecules in the same chain causes the effective interaction between dendrimer
generations.
Namely, the interaction between molecules in the chain linking the site
$i$ with the site $i+1$ makes up the interaction coefficient $V_{ij}$.
So, $V_{ij}$
actually is the effective interaction, that will be calculated in
the section III.

We apply here the " real space" Green's function method developed in [13, 14].
The double-time Green's functions are writtten as

\begin{equation}
g_{ij}=<<a_i(t), a^+_j(0)>> = -i\Theta(t)<[a_i(t),a_j^+(0)]>
\end{equation}

\noindent
where [,] is a commutator, a(t) is the Heisenberg representation of the operator a,

\begin{equation}
a(t) = e^{iHt}ae^{-iHt}
\end{equation}

\noindent
$\Theta(t)$ is Heaviside function, $<...>$ is the thermal average over a grand canonical
ensemble.Using the Fourier transformation to transfer the time Green's function
to the energy variable, we have

\begin{equation}
g_{ij}(t) = \int_{-\infty}^{\infty}{g_{ij}(E)e^{-iEt/\hbar }dE}
\end{equation}

\noindent
By using the Green function

\begin{equation}
G_{ij}(E) = \frac{2\pi}{\hbar}g_{ij}(E)
\end{equation}

\noindent
and the Heisenberg equation of motion

\begin{equation}
i\hbar\dot{a}_j = {\cal E}_ja_j + \sum_{i}V_{ji}a_{i}
\end{equation}

\noindent
we have Schwinger-Dyson equations for the Green's functions for our system

\begin{equation}
(E-{\cal E}_i)G_{ij}(E) -\sum_{i,k}V_{ik}G_{kj}(E) = \delta_{ij}
\end{equation}

\noindent
Since the system consists of both Wannier and Frenkel excitons, the Green funtions
we consider here are the Green's function of hybridized state and E is the energy
of the new excitation.

In the next section we will calculate the effective interaction
cofficients $V_{ij}$, then in the following sections we will
apply the Schwinger equations (7) to solve for our system.

\vspace{0.5cm}

{\bf III. The effective interaction coefficients}

\vspace{0.5cm}

As it has been said in the section II, the interaction between different
generations, called the interaction at a macroscopic scale, has been made
from all the interactions between nearest neighboring molecules, called the interaction
at the microscopic scale.
For example, we consider a chain as in the Fig.2a. We want to calculate the
interaction between the quantum dot at the center, which we can consider
as the zero-generation, and the molecule at the end of one of the first
molecular chains, which is the first generation.
In the chain there is a number of molecules, so the quantum
dot and the molecule finishing the first generation may be far apart and do
not interact directly with each other if we assume only nearest neighbor
interaction.
So the exciton in the quantum dot interacts with the exciton in the first
molecule of the first chain, the first molecule interacts with the second
molecule, and so on, the next to the last molecule interacts with the last
molecule of the chain.
Actually in this process the exciton in the quantum dot interacts with the
exciton in the last molecule of the first chain indirectly through the effective
"superexchange" interaction. The problem is very similar to the problem of
electron transfer in a chain with one impurity [15]. In this section we use the Green's function and
renormalization approach [15, 16 ] to calculate the effective interaction.

As in [15], for the linear chain we consider the Huckel Hamiltonian:
\begin{equation}
H= \sum_i{\epsilon_{\alpha}a^+_{\alpha}a_{\alpha}} 
+\sum_{\alpha,\beta}v_{\alpha,\beta}(a^+_{\alpha}a_{\beta} + a^+_{\beta,a_{\alpha})} 
\end{equation}

\noindent
$a^+_{\alpha}$, ( $a_{\alpha}$) are the creation (annihilation) exciton operator at the site 
${\alpha}$.
$v_{\alpha,\beta}$ is the interaction of excitons at site ${\alpha}$ and site ${\beta}$.
Here we assume only nearest neighbor interaction, so ${\beta} ={\alpha}\pm 1$.
${\alpha}$  numbers sites of the linear chain, namely if our linear chain consists of
n molecules, the quantum dot lies at the site ${\alpha}=0$, and the molecules lie
at sites 1,2,...n.
The nearest neighbor interaction between excitons in adjacent sites
is $v_{\alpha,\alpha +1}$, where $v_{01}$ is the interaction between the Wannier
exciton in the quantum dot and the Frenkel exciton in the organic molecule,
and all other $v_{12}, v_{23},..., v_{n-1,n}$ are the interaction between
the Frenkel excitons in the molecules.
Then $v_{0n}$ will be the effective interaction between the quantum dot
and the first generation in the "macroscopic scale". Then we can replace the chain
by the quantum dot and the last molecule of the chain interacting with each other
through  $v_{0n}$.
The Dyson's equation for this Hamiltonian [15]:

\begin{equation}
Eg_{\alpha \beta}=\sum_kH_{\alpha k}g_{k \beta}
\end{equation}

\noindent
For our chain (Fig.2a) this equation can be written as:

\begin{eqnarray}
(E-\epsilon_0)g_{00} & = & 1 +v_{01}g_{10} \nonumber \\
(E-\epsilon_1)g_{10} & = & v_{10}g_{00} + v_{12}g_{20} \nonumber \\
(E-\epsilon_2)g_{20} & = & v_{21}g_{10} + v_{23}g_{30} \nonumber \\
& ... & \nonumber \\
(E-\epsilon_{n-1})g_{n-1,0} & = & v_{n-1,n-21}g_{n-2,0} + v_{n-1,n}g_{n0}
\nonumber \\
(E-\epsilon_n)g_{n,0} & = & v_{n,n-1}g_{n-1,0} 
\end{eqnarray}

\noindent
Here $\epsilon_0 = \epsilon_W$ is the energy of the Wannier exciton in the quantum dot,
$\epsilon_1, \epsilon_2,...\epsilon_n$ are the energies of the Frenkel excitons in the molecules.
since we consider the dendrimetic chain of identical molecules, so we can assume
all $\epsilon_1, \epsilon_2,...\epsilon_n = \epsilon_F$.
Also, except for the dot-molecule Wannier exciton-Frenkel exciton interaction $v_{01} = 
v_{WF}$, all other $v_{i,i+1}=v$ are the same.
Then the system (10) becomes:

\begin{eqnarray}
(E-\epsilon_W)g_{00} & = & 1 +v_{0}g_{10} \nonumber \\
(E-\epsilon_F)g_{10} & = & v_{0}g_{00} + vg_{20} \nonumber \\
(E-\epsilon_F)g_{20} & = & vg_{10} + vg_{30} \nonumber \\
& ... & \nonumber \\
(E-\epsilon_F)g_{n-1,0} & = & vg_{n-2,0} + vg_{n0}
\nonumber \\
(E-\epsilon_n)g_{n,0} & = & vg_{n-1,0} 
\end{eqnarray}

\noindent
Outside the defect region ($i>2$), the Green's functions of neighboring sites are supposed 
to be related to each other by some transfer function T 
with $G_{i+1,0} = TG_{i,0}$. By replacing this relation into the third and the following
equations in the system (10), we obtain for the transfer function T:

\begin{equation}
T =\frac{E-\epsilon_F\pm \sqrt{(E-\epsilon_F)^2-4v^2}}{2v}
\end{equation}

\noindent
and for the Green's functions

\begin{eqnarray}
g_{00} & = & \frac{1}{E-\epsilon_W -\frac{v^2_0}{E-\epsilon_F - vT}}\nonumber \\
g_{10} & = & \frac{v_0}{E-\epsilon_F-vT}g_{00}
\end{eqnarray}

\noindent
and
\begin{equation}
g_{n0} = \frac{v_0}{E-vT}T^{n-1}g_{00}
\end{equation}

\noindent
from here we can have the "effective" interaction between site 0 and site n, or
between the exciton in the quantum dot and the Frenkel exciton of the
first dendrimeric generation:

\begin{equation}
v_{0n} =  \frac{v_0}{E-vT}T^{n-1} = V_{WF}
\end{equation}

\noindent
And from the Green's function (13) we can have for the energy E of the mixed state
in this chain

\begin{equation}
E = \epsilon_W +\frac{v^2_0}{E-\epsilon_F - vT}
\end{equation}

\noindent

In this fashion we already obtain the "effective" interaction between exciton in
the quantum dot and the exciton of the first generation. Similarly
we have the effective interaction between the Frenkel excitons in
the neighboring generations like between the first and the second
generations or between the second and the third etc...Like $V_{WF}$,
this effective interaction $V_{FF}$ is also made up of all the nearest
neighbor interactions between the excitons of the molecules in the chain
connected the last molecules of the two generations (Fig.2b).
Similar to (15) we have

\begin{equation}
V_{FF} =  \frac{v}{E-vT}T^{n-1} 
\end{equation}

\noindent
and the energy in the chain
\begin{equation}
E = \epsilon_F +\frac{v^2}{E-\epsilon_F - vT}
\end{equation}

\noindent
Again, n is the number of molecules in the chain.

We note here that the "effective interaction" depends on the energies of
excitons, the nearest neighbor interaction coefficients, and also on the
number of molecules in the chain.

Now that we have all the needed interaction coefficients for the
Schwinger-Dyson equation (7), we can replace the molecular chains by localized sites with 
effective interaction and we will apply this equation for our particular quantum dot-dendrimer 
system. \vspace{0.5cm}

{\bf IV.Diagram Techniques for Green's Function in Orbital Representation}

\vspace{0.5cm}

To solve the Schwinger-Dyson equations for the dendrimer system with
many generations, it is convenient to use the "electron transfer
graph" method developed in [13,14], where the method is used to
study electron transfer between localized sites. Then for the Schwinger-Dyson
equation (7),
every site corresponds to a graph vertex. A non-diagonal term $V_{jk}$
of the Hamiltonian corresponds to an oriented edge originating at vertex j
and ending at vertex k with the value of the edge equal to the interaction integral
$V_{jk}$. Diagonal terms $E - {\cal E}_{ii}$ of the Hamiltonian corresponds
to a loop attached to the vertex i and the value of the loop equals 
$W_{ii}=(E-{\cal E}_i)$.

The details of the method and the proof of some formulae can be found in [13]. 
In this section we just write briefly about definitions and terminology, the main method and  
results of the authors[13] which we will use in the following sections.

In this diagram technique a path is a sequence of graph edges with succesive edges 
originating at the point where the previous one ends, and a cycle is a path with the last edge 
ending at the point where the first one originates. A length of the path is the number of 
edges in it, where $P_{ii} =1$.  The value O of the cycle is the product of the values of all 
edges in the cycle, with the sign being negative except for the loop for any cycle with 
length more than 1. The cyclic term is a set of cycles which have no common vertice and pass 
through every vertex. The value of  cyclic term $ {\bf O}$ is a product of the value of all 
cycles of the cyclic term and the cyclic value of the graph $\Theta$ is the sum of all the 
cyclic terms.

For illustration, if the chain of sites is three sites as in Fig.3b,
the cyclic term and the cyclic value of the graph will be as in the Fig.3c,
or [13]:
\begin{eqnarray}
\Theta_{1,3} & = & {\bf 0_1} + {\bf 0_2} + {\bf 0_3} \nonumber \\
             & = & (E-{\cal E}_1)(E-{\cal E}_2)(E-{\cal E}_3)
             - V_{12}V_{21}(E-{\cal E}_3) - V_{23}V_{32}(E-{\cal E}_1)
\end{eqnarray}

\noindent
For a chain of sites as in Fig.3a it is shown in [13] that the Green's
function $G_{ij}$ is equal to

\begin{equation}
G_{ij} = \frac{\Theta_{1,i-1}P_{ij}\Theta_{j+1,n}}{\sum{O^{\{k\}}}}
\end{equation}

\noindent
where $P_{ij}$ is the product of the edges along the pathway from vertex 
i to vertex j
\begin{equation}
P_{ij} = V_{i,i+1}V_{i+1,i+2}...V_{j-1,j}
\end{equation}

\noindent
$\Theta=\sum{O^{\{k\}}} (i,j)$ is the cyclic value of the graph and
$O^{\{k\}}_{ij}$ is the cyclic term.

Explicitly the Green functions have the form:
\begin{eqnarray}
G_{11} & = & \frac{1}{\Theta}[(E-{\cal E}_2)(E-{\cal E}_3) - V_{12}V_{21}]\nonumber\\
G_{12} & = & \frac{1}{\Theta}V_{12}(E-{\cal E}_3) = G_{21} \nonumber \\
G_{13} & = & \frac{1}{\Theta}V_{12}V_{23} = G_{31} \nonumber \\
G_{22} & = & \frac{1}{\Theta}(E-{\cal E}_1)(E-{\cal E}_3) \nonumber \\
G_{23} & = & \frac{1}{\Theta}(E-{\cal E}_1)V_{23} = G_{32} \nonumber\\
G_{33} & = & \frac{1}{\Theta}[(E-{\cal E}_1)(E-{\cal E}_2) - V_{23}V_{32}]
\end{eqnarray}

\noindent
For the case when there are sites of the sidegroup attached to some site i of the
path (Fig.4a) then all the graph of the main chain remains the same, only the value of
the loop at site i is different. In this case it was proved in [13] that
the diagonal element of the Green's
function of side-graph can be expressed through the Green's function of subgraph
in the continued fraction representation.

The value of the loop with the sidegroup could be expanded into continuous
fraction of loops values of extended side graph $\tilde{\gamma}$-the subgraph
of the graph. In another word, instead of the original loop value
$G^{-1}_i = W_{ii} = E-{\cal E}_i$ the loop value of the site with sidegroup becomes:

\begin{eqnarray}
\tilde{G}^{-1}_i = \tilde{W}_{ii} & = & (E-{\cal E}_i) -\sum_k{\frac{V^2_i,k(i)}
{\tilde{\Theta}_{k(i)}/\Theta_{k(i)}}}\nonumber \\
       & = & (E-{\cal E}_i) -\sum_k{V^2_{i,k(i)}G(\tilde{\gamma}_k(i))_{kk}}
\end{eqnarray}

\noindent
For illustration for the site chain in the Fig.4b the value of the loop at
site 2 has the form:

\begin{equation}
\tilde{W}_{22} = E-{\cal E}_2 - \frac{V_{2a}V_{a2}}{E-{\cal E}_a -\frac{V_{ab}V_{ba}}
{E-{\cal E}_b}}
\end{equation}

\noindent
by this continuous fraction representation we can count all the tree structure
attached to any site of the chain.

Now we come back to consider our model of quantum dot-dendrimer system in the
section II and apply this method to calculate the Green's function matrix elements
for the hybridization excitation.

\vspace{0.5cm}

{\bf V. The quantum dot- dendrimeric "slice" }

\vspace{0.5cm}
As the first model for quantum dot-dendrimeric system, we consider the model
in Fig.5a where a quantum dot is attached to a chain of molecules.
The last molecule of this chain in its turn is connected to two
chains, the last molecules of each is again connected with two other
chain and so on. The number of branching points along one chain from the center to the
terminal point is the generation number of the dendrimer. This model is very similar to the 
dendritic slice attached to the dot in several experiments[4,5].

Using the diagram with continous fraction representation  we can consider the
slice as a linear chain with sidegroups attached to each site.
Then within this approximation where we neglect some symmetry in this structure the Fig.5a 
can be considered as the Fig. 5b.

For the quantum dot-dendrimer slice of 2 generations (Fig.5c) we have:

\begin{equation}
\Theta = (E - {\cal E}_W)(E - {\cal E}_F -\frac{V^2_{FF}}{E - {\cal E}_F})(E-{\cal E}_F)
-V^2_{WF}(E - {\cal E}_F) - V^2_{FF}(E -{\cal E}_W)
\end{equation}

\noindent
and Green's functions:
\begin{eqnarray}
G_{00} & = & \frac{(E-{\cal E}_F-\frac{V^2_{FF}}{E-{\cal E}_F})(E-{\cal E}_F)-
V_{WF}V_{FW}}{\Theta}\nonumber \\
G_{01} & = & \frac{V_{WF}(E-{\cal E}_F) }{\Theta} \nonumber \\
G_{02} & = & \frac{V_{WF}V_{FF}}{\Theta}  \nonumber\\
G_{12} & = & \frac{(E-{\cal E}_W)V_{FF}}{\Theta}
\end{eqnarray}

\noindent
and for the system with 3 generations of dendrimeric chains (Fig.5d):

\begin{equation}
\Theta = (E-{\cal E}_W)\frac{1}{G_1G_2G_3} +
V_{WF}V^3_{FF} - \frac{V^2_{WF}}{G_2G_1}
 -\frac{V^2_{FF}(E-{\cal E}_W)}{G_3} - \frac{(E-{\cal E}_W)V^2_{FF}}{G_1}
\end{equation}

\noindent
with the loop values:

\begin{eqnarray}
\frac{1}{G_3} & = & E-{\cal E}_F \nonumber \\
\frac{1}{G_2} & = & E- {\cal E}_F - \frac{V^2_{FF}}{E- {\cal E}_F}\nonumber\\
\frac{1}{G_1} & = & E- {\cal E}_F -2\frac{V^2_{FF}}{E-{\cal E}_F}
\end{eqnarray}

\noindent
And similarly to (26) we also calculated the Green's function matrix
elements $G_{01}, G_{12}...G_{33}$.

In this way we always can calculate Green's function matrix for the cases
of more generations.

Note here that the energy of the new excitation in the system, i.e. the energy
of the Wannier-Frenkel hybrid exciton is the zero condition for $\Theta$ (25), (27).
The energy and also the Green functions depend on the energies of both the Wannier and the Frenkel excitons, on
the effective interactions coefficient, and the structure of the
dendrimer branches. So the geometry of the dendrimer, the number of the
branches are very important. To change the geometry of the quantum dot- dendrimer
system, we will change the energy of the hybrid exciton and then the optical processes
of the system.
\vspace{0.5cm}

{\bf VI. The quantum dot-dendron ligand}

\vspace{0.5cm}
In recent time several series of semiconductor nanocrystal- organic dendron ligands, an 
objective which has hyperbranched organic molecules, have been designed and synthesized [5].
The priority of the dendron ligand is perfect symmetry with the dot at the center, and also the 
closely packed ligand shell.

We can apply the diagram method to study the hybrid excitation for this model dendron. We
will consider in this section the ligand  model we mentioned in section II.
A quantum dot is attached with three branches, then each terminal molecule
is attached to other two branches, and so on (Fig.1).
At first we consider the first generation system with the quantum dot at the
cente attached to 3 chains of molecule (Fig. 6a). Consider this system
like a linear chain of the chain 1, the dot and the chain 3. the chain 2 then is
considered as the chain 2' attaching to the dot. For this case we have:

\begin{eqnarray}
G_{11} & = & \frac{1}{\Theta}[(E-\tilde{{\cal E}_2})(E-{\cal E}_3) - V_{12}V_{21}]\nonumber\\
G_{12} & = & \frac{1}{\Theta}V_{12}(E-{\cal E}_3) = G_{21} \nonumber \\
G_{13} & = & \frac{1}{\Theta}V_{12}V_{23} = G_{31} \nonumber \\
G_{22} & = & \frac{1}{\Theta}(E-{\cal E}_1)(E-{\cal E}_3) \nonumber \\
G_{23} & = & \frac{1}{\Theta}(E-{\cal E}_1)V_{23} = G_{32} \nonumber\\
G_{33} & = & \frac{1}{\Theta}[(E-{\cal E}_1)(E-\tilde{{\cal E}_2}) - V_{23}V_{32}]
\end{eqnarray}

\noindent
and
\begin{equation}
\Theta = (E-{\cal E}_1)(E-\tilde{{\cal E}_2})(E-{\cal E}_3)
             - V_{12}V_{21}(E-{\cal E}_3) - V_{23}V_{32}(E-{\cal E}_1)
\end{equation}

\noindent
where the  value of the loop at site 2 (the dot) is including the side graph
\begin{equation}
\tilde{{\cal E}_2} = {\cal E}_2-\frac{V_{22'}V_{2'2}}{E-{\cal E}_{2'}}
\end{equation}

\noindent
In the language of Wannier and Frenkel exciton, then we have for the
Green functions of the system:

\begin{eqnarray}
G_{11} & = & \frac{1}{\Theta}[(E-{\cal E}_W-\frac{V_{WF}V_{FW}}{E-{\cal E}_F} 
(E-{\cal E}_F) - V_{FW}V_{WF}]\nonumber\\
G_{12} & = & \frac{1}{\Theta}V_{FW}(E-{\cal E}_F) = G_{21} \nonumber \\
G_{13} & = & \frac{1}{\Theta}V_{FW}V_{WF} = G_{31} \nonumber \\
G_{22} & = & \frac{1}{\Theta}(E-{\cal E}_F)^2 \nonumber \\
G_{23} & = & \frac{1}{\Theta}(E-{\cal E}_F)V_{WF} = G_{32} \nonumber\\
G_{33} & = & \frac{1}{\Theta}[(E-{\cal E}_F)(E-{\cal E}_W-
\frac{V_{WF}V_{FW}}{E-{\cal E}_F} - V_{FW}V_{WF}]
\end{eqnarray}

\noindent
and
\begin{equation}
\Theta  =  (E-{\cal E}_F)^2(E-{\cal E}_W-\frac{V_{WF}V_{FW}}{E-{\cal E}_F})
             - V_{FW}V_{WF}(E-{\cal E}_F) - V_{WF}V_{FW}(E-{\cal E}_F)
\end{equation}

\noindent
For the two generation quantum dot-dendron, we can consider the dendron as
linear chain of one generation  with sidegroup attached to each of the terminal
molecules (Fig.6b).
Then the loop value ${\cal E}_1,{\cal E}_2,{\cal E}_3$ become

\begin{eqnarray}
\tilde{{\cal E}_1} & = &{\cal E}_1 -\frac{V_{1,11}V_{11,1}}{E-{\cal E}_{11}}
-\frac{V_{1,12}V_{12,1}}{E-{\cal E}_{12}}\nonumber\\
\tilde{{\cal E}_3} & = &{\cal E}_3 -\frac{V_{3,31}V_{31,3}}{E-{\cal E}_{31}}
-\frac{V_{3,32}V_{32,3}}{E-{\cal E}_{32}} \nonumber\\
\tilde{{\cal E}_1}' & = & {\cal E}'_1 -\frac{V_{2',2'1}V_{2'1,2'}}{E-{\cal E}_{2'1}}
-\frac{V_{2,2'2}V_{2'2,2'}}{E-{\cal E}_{2'2}}
\end{eqnarray}

\noindent
Replace  these values into the formulae for the Green function (29), we have Green's
functions for hybrid exciton in the system  of 2 generation of dendrimer.

Continue this process of sidegroup graph attached to those three chains of the first
generation, practically we can go farther to any generation just by changing the loop values
in (29). Since every branching point is attached to two branches, in going from n generation to n+1 
generation, the value of the loop i just needs addition by the value  
of the two branches of n+1 generation attached to the generation n of the loop:
\begin{equation}
\tilde{{\cal E}^{n+1}_i} = \tilde{{\cal E}^n_i} - \frac{V_{in;in+1,1}V_{in+1,1;in}}{E - {\cal 
E}_{i,n+1,1}} - \frac{V_{in;in+1,2}V_{in+1,2;in}}{E - {\cal E}_{i,n+1,2}}
\end{equation}

\noindent
where 1 and 2 denote the two branches of the n+1 generation.  From here, if we know the energy and 
Green's function of the dendrimer with n generations, we always can go to the dendrimer with n+1 
generations. Because of the continuous fraction, from some generation, the effects of 
 n+1 generation is small and can be treated as some small perturbations.

The energy of the Wannier-Frenkel hybrid exciton again is the pole of the Green functions, or in another 
words, is obtained from the condition for the zeros of the Green's function determinator 
$\Theta$. For example 
for the quantum dot-dendron ligand of one generation the equation (33) gives us:
\begin{equation}
E^{W-F} = \frac{1}{2}[{\cal E}_F + {\cal E}_W \pm \sqrt{({\cal E}_F -{\cal E}_W)^2 + 8V_{WF}V_{FW}}]
\end{equation}

\noindent
and for the quantum dot-dendron ligand of two generation we have for energy of the hybrid exciton:
\begin{eqnarray}
E^{W_F} & = & \frac{1}{2}({\cal E}_F +{\cal E}_W +\frac{4V^2_{FF}}{E-{\cal E}F} - 
\frac{2V^2_{WF}}{E-{\cal E}W} \pm [ {\cal E}^2_F + {\cal E}^2_W + 6{\cal E}_W{\cal E}_F \nonumber \\
&& + \frac{16 V^4_{WF}}{(E -{\cal E}_F)^2} - \frac{16 V^4_{FF}}{(E -{\cal E}_F)^2} + \frac{24 
V^2_{FF}V^2_{WF} }{(E -{\cal E}_F)^2} + 8V^2_{WF}]^{1/2}])
\end{eqnarray}

\noindent
we mention here that using the continuous fractional diagram technique, each time when we 
have more 
generations, the values of $ \frac{2V^2_{FF}}{E-{\cal E}F}$ are added to each branching point. It means 
when one more generation is added, in the  equation (33) only the value of E will change for $ 
\frac{2V^2_{FF}}{E-{\cal E}F}$ time some integers.
Then for n large when it is difficult to solve (33) analitically, the self-consistent method 
is used to solve 
the equation, so for quantum dot-dendrom ligands of any generation we always have expressions 
for the
energy of the hybrid exciton.

For a simply illustrative numerical calculation we use CdSe quantum dot with pentacene 
dendrimer. If the radius of the dot is 
3.8 nm $E^W = 2.1 ev$, $ E^F = 1.5 ev$,$ \mu^F = 5D, \mu^W = 10 D$. 
Assume there are 4 molecules in one chain  then  there are two values of hybrid exciton $ E 
^{W_F} = 2.1035 ev $ and $ E ^{W_F} = 1.4996 ev$. It means 
there are two hybrid exciton values close to  the Wannier and Frenkel excitons, with a 
shift of about 0.4 mev. If we 
use tetracene dendrimer with Frenkel exciton energy $E^F= 2 ev$, closer to energy of 
the Wannier exciton in the 
quantum dot, then we get two values for the hybrid exciton at $ E ^{W_F} = 2.1019 ev $ 
and $ E ^{W_F} = 1.9981 ev$, or the shift is about 1.95 mev. 
So if we choose the size of dot and material of dendrimer
so that the energies of the Wannier and Frenkel excitons are close, the effect of 
hybridization is larger. 
The structure of the dendrimer (number of molecule in one branch, for instance), also decides 
the hybridization.

Green's functions are quantities to charaterize a many-particle system, so obtaining Green's 
functions gives a lot of information of the system. Since any particular observable can be calculated 
using the Green's function relations such as
\begin{equation}
<O> = i\int{dxdx'O(xx')G(xt|x't+)}
\end{equation}

\noindent
we can use the real-time components of the Green's functions (29) to calculate physical quantities 
of the systems.

 {\bf Summary}

In this paper we propose a model to study the organic-semiconductor hybrid exciton in  quantum 
dot-dendrimer system. The energy of the hybrid exciton as well as the Green's function matrix 
elements  for different quantum dot-dendrimer systems have been calculated. With the Green's 
functions the optical properties and optical processes are expected to be obtained. By changing 
structure and parameter of the systems, the optical properties can be controlledas interested. 
The use of these Green Functions to calculate optical and other response functions will be 
discussed elsewhere (see Ref. 17).

 {\bf Acknowledgements}

We acknowledge support in part from NYSTAR contract $ N^o0000067$. We would also like to thank 
Prof. J. Malinski and Prof. V. Balogh-Nair for useful discussions.

\newpage
{\bf FIGURE CAPTIONS}
\vspace{2cm}

\noindent
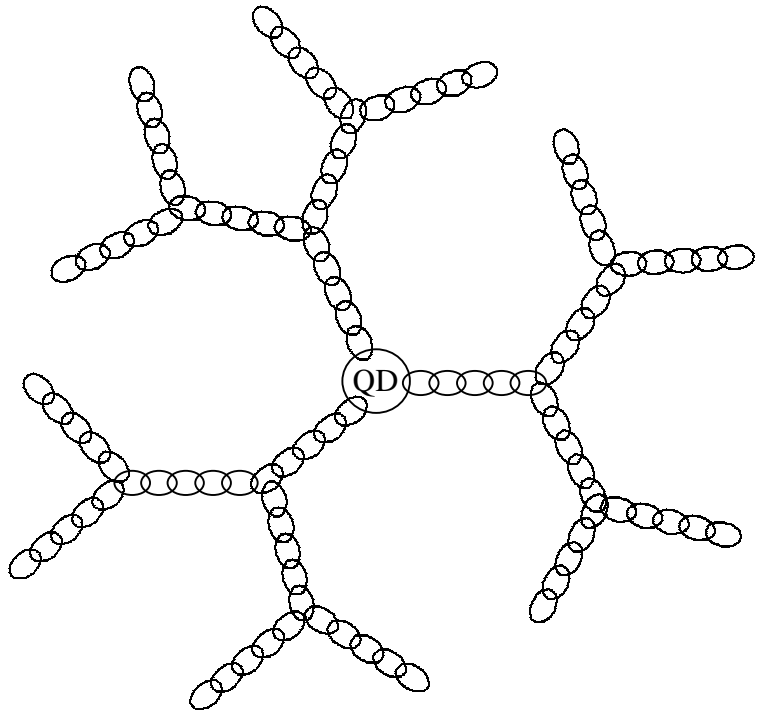
{\bf Fig.1}. A Quantum Dot- Dendron Ligand Model . 
\vspace{0.7cm}

\noindent
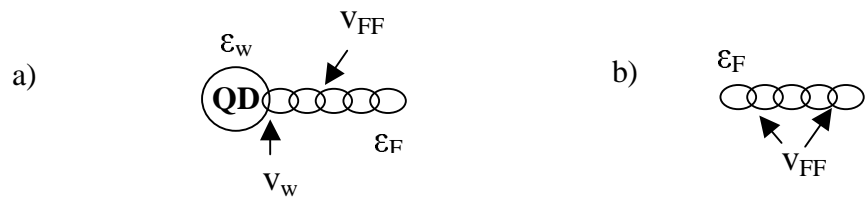
{\bf Fig.2} a) A quantum dot-molecule chain. b) A molecule-molecule chain.\\
\vspace{0.7cm}

\noindent
{\bf Fig.3}. a)Linear chain of localized sites. b)Linear chain with three vertices.
c)Cyclic value of the graph for linear chain of three vertices\\
\vspace{0.7cm}

\noindent
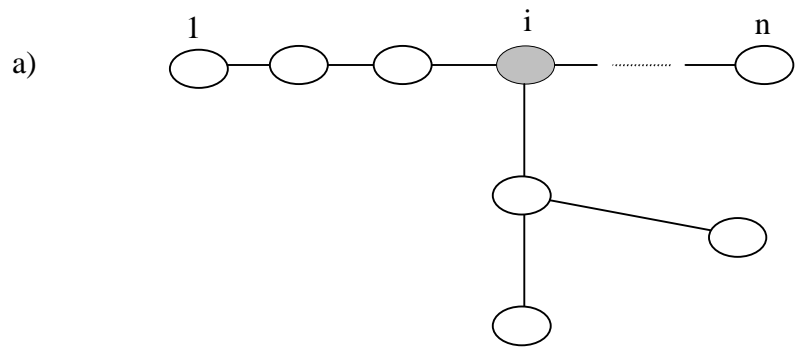
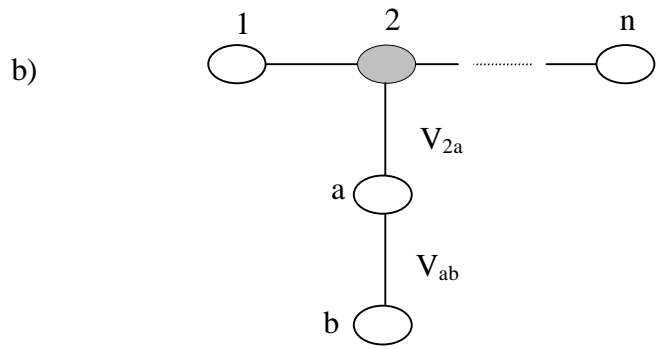
{\bf Fig.4}. a) A chain with a side group attached to the site i. b)A graph of a side group. 
\vspace{0.7cm}

\noindent
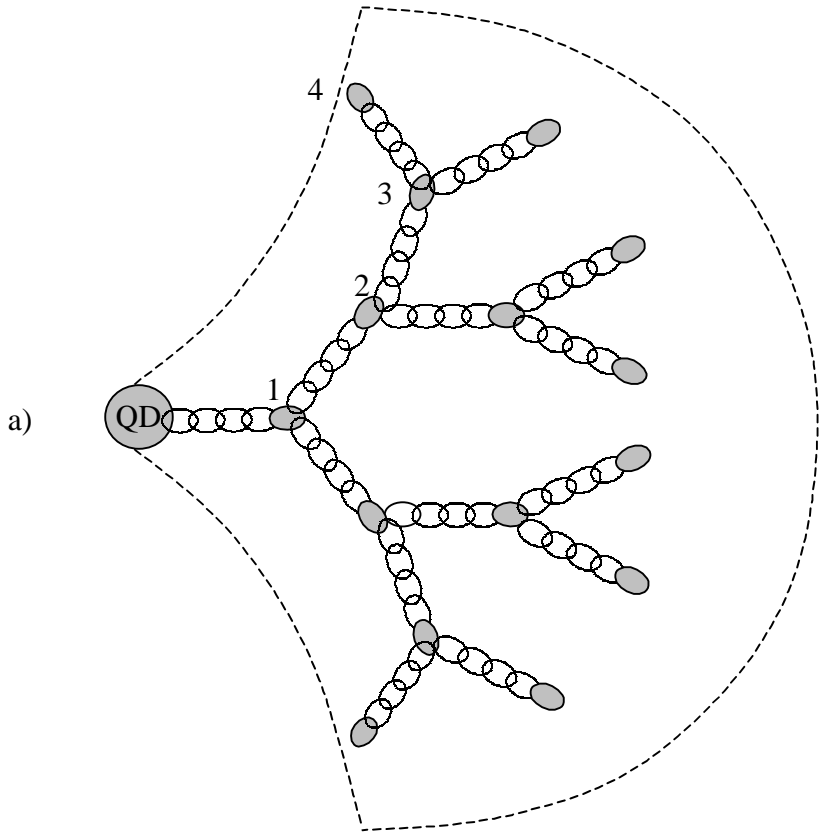
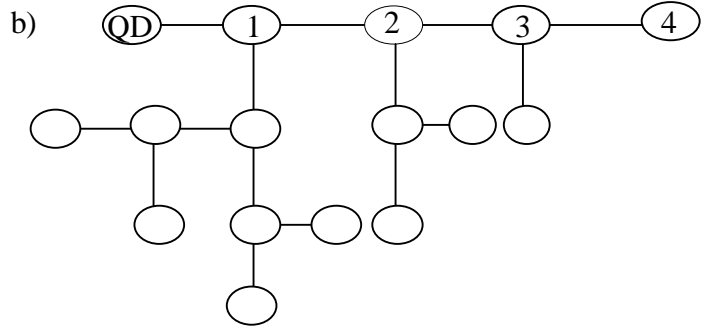
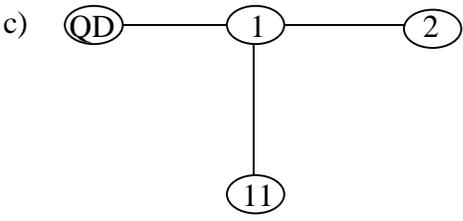
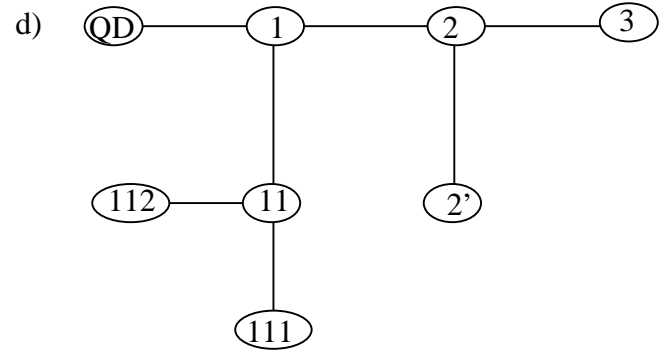
{\bf Fig. 5} a)A quantum dot-dendrimer slice. b) A QD-dendrimer slice as a side group graph. c) A 
slice of two generations. d)A slice of three generations.\\ 
\vspace{0.7cm}

\noindent
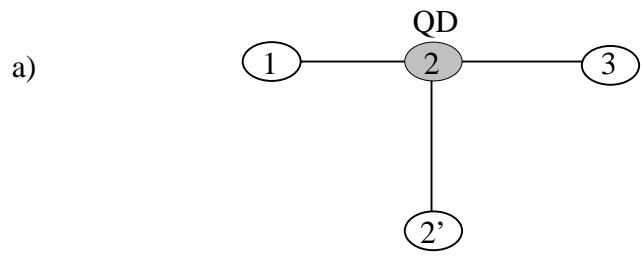
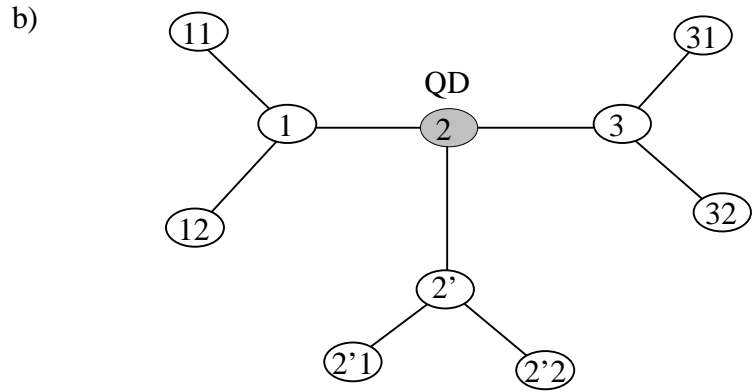
{\bf Fig.6}. a)A QD- dendron ligand with one generation. b)A QD- dendron ligand with two or more 
generations. 

\end{document}